\documentclass[
	conference,
	10pt,
    a4paper
]{IEEEtran}

\IEEEoverridecommandlockouts % This is to make the \thanks{} command visible in conference mode
\makeatletter
\def\markboth#1#2{\def\leftmark{\@IEEEcompsoconly{\sffamily}\MakeUppercase{\protect#1}}%
\def\rightmark{\@IEEEcompsoconly{\sffamily}\MakeUppercase{\protect#2}}}
\makeatother

\usepackage[keeplastbox]{flushend}
\usepackage[amssymb]{SIunits}
\usepackage[cmex10]{amsmath} 
\usepackage{amssymb, amsthm}
\usepackage{bbm} % indicator function \mathbbm{1}
\usepackage{bm}
\usepackage{xcolor}
\usepackage{cite}
\usepackage{graphicx}
\usepackage[caption=false, font=footnotesize]{subfig}
\usepackage{url}
\usepackage{mathtools}

\usepackage{xspace} % for the ``space after newcommand'' problem
\usepackage{tikz,pgfplots}

\usepackage{booktabs} % for \toprule, \midrule and \bottomrule macros

\usepackage{array}
\newcolumntype{L}[1]{>{\raggedright\let\newline\\\arraybackslash\hspace{0pt}}m{#1}}
\newcolumntype{C}[1]{>{\centering\let\newline\\\arraybackslash\hspace{0pt}}m{#1}}
\newcolumntype{R}[1]{>{\raggedleft\let\newline\\\arraybackslash\hspace{0pt}}m{#1}}

\usepackage[
	shortcuts,
	nonumberlist,	% don't show page numbers
	acronym, % use acronyms
	hyperfirst=true,
	section=section, % section-Ebene
	order=letter,
	numberedsection=nolabel % false / nolabel / autolabel
]{glossaries}

\theoremstyle{definition}

\theoremstyle{plain}

\theoremstyle{remark} % subsequently defined environments use this style

\newcommand{\transpose}{\top}

\newcommand{\vect}[1]{\ensuremath{\boldsymbol{#1}}}
\newcommand{\mat}[1]{\ensuremath{\boldsymbol{#1}}}

\newcommand{\imag}{\jmath}

\newcommand{\StepSize}{\ensuremath{\delta}}
\newcommand{\NumSteps}{\ensuremath{M}}

\newif\ifshow
\showtrue
\newcommand{\abbr}[1]{{#1}}				

\makeatletter
\let\aclOLD=\acl
\renewcommand{\acl}[1]{%
  \begingroup    
  \let\@@underline=\relax
  \aclOLD{#1}%
  \endgroup
}
\makeatother

\newcommand{\NewA}[3]{
	\newacronym{#1}{#2}{#3}
}

\NewA{af}{AF}{\abbr{a}mplify-and-\abbr{f}orward}
\NewA{apsk}{APSK}{\abbr{a}mplitude \abbr{p}hase-\abbr{s}hift \abbr{k}eying}
\NewA{ask}{ASK}{\abbr{a}mplitude-\abbr{s}hift \abbr{k}eying}
\NewA{ase}{ASE}{\abbr{a}mplified \abbr{s}pontaneous \abbr{e}mission}
\NewA{awgn}{AWGN}{\abbr{a}dditive \abbr{w}hite \abbr{G}aussian \abbr{n}oise}
\NewA{biawgn}{BI-AWGN}{binary-input \abbr{a}dditive \abbr{w}hite \abbr{G}aussian \abbr{n}oise}
\NewA{bep}{BEP}{\abbr{b}it \abbr{e}rror \abbr{p}robability}
\NewA{cep}{CEP}{\abbr{c}odeword \abbr{e}rror \abbr{p}robability}
\NewA{ber}{BER}{\abbr{b}it \abbr{e}rror \abbr{r}ate}
\NewA{qap}{QAP}{\abbr{q}uadratic \abbr{a}assignment \abbr{p}roblem}
\NewA{bicm}{BICM}{\abbr{b}it-\abbr{i}nterleaved \abbr{c}oded \abbr{m}odulation}				
\NewA{cm}{CM}{\abbr{c}oded \abbr{m}odulation}				
\NewA{qpsk}{QPSK}{quadrature phase-shift keying}				
\NewA{mlcm}{MLCM}{multilevel coded modulation}				
\NewA{bpsk}{BPSK}{\abbr{b}inary \abbr{p}hase-\abbr{s}hift \abbr{k}eying}				
\NewA{bsc}{BSC}{\abbr{b}inary \abbr{s}ymmetric \abbr{c}hannel}				
\NewA{brgc}{BRGC}{\abbr{b}inary \abbr{r}eflected \abbr{G}ray \abbr{c}ode}				
\NewA{cf}{CF}{\abbr{c}haracteristic \abbr{f}unction}
\NewA{csit}{CSIT}{\abbr{c}annnel \abbr{s}tate \abbr{i}nformation at the \abbr{transmitter}}
\NewA{csi}{CSI}{\abbr{c}annnel \abbr{s}tate \abbr{i}nformation}
\NewA{df}{DF}{\abbr{d}ecode-and-\abbr{f}orward}
\NewA{fd}{FD}{\abbr{f}ull-\abbr{d}uplex}
\NewA{fft}{FFT}{\abbr{f}ast \abbr{F}ourier \abbr{t}ransform}
\NewA{iid}{IID}{independent and \abbr{i}dentically \abbr{d}istributed}
\NewA{isi}{ISI}{\abbr{i}nter\abbr{s}ymbol \abbr{i}nterference}
\NewA{lb}{LB}{\abbr{l}ower \abbr{b}ound}
\NewA{map}{MAP}{\abbr{m}aximum \abbr{a} \abbr{p}osteriori}
\NewA{mf}{MF}{\abbr{m}odulo-and-\abbr{f}orward}
\NewA{mlan}{MLAN}{\abbr{m}odulo-\abbr{l}attice \abbr{a}dditive \abbr{n}oise}
\NewA{ml}{ML}{\abbr{m}aximum \abbr{l}ikelihood}
\NewA{mmse}{MMSE}{\abbr{m}inimum \abbr{m}ean \abbr{s}quare \abbr{e}rror}
\NewA{nlpc}{NLPC}{\abbr{n}onlinear \abbr{p}hase \abbr{c}ompensation}
\NewA{nlpn}{NLPN}{\abbr{n}onlinear \abbr{p}hase \abbr{n}oise}
\NewA{nc}{NC}{\abbr{n}etwork \abbr{c}oding}
\NewA{ofmd}{OFDM}{\abbr{o}rthogonal \abbr{f}requency-\abbr{d}ivision \abbr{m}ultiplexing}			
\NewA{dp}{DP}{\abbr{d}ual-\abbr{p}olarization}
\NewA{pam}{PAM}{\abbr{p}ulse \abbr{a}mplitude \abbr{m}odulation}
\NewA{pdf}{PDF}{\abbr{p}robability \abbr{d}ensity \abbr{f}unction}
\NewA{plnc}{PNC}{\abbr{p}hysical-layer \abbr{n}etwork \abbr{c}oding}
\NewA{psk}{PSK}{\abbr{p}hase-\abbr{s}hift \abbr{k}eying}
\NewA{pmd}{PMD}{\abbr{p}olarization \abbr{m}ode \abbr{d}ispersion}
\NewA{pm}{PM}{\abbr{p}olarization-\abbr{m}ultiplexed}
\NewA{pdm}{PDM}{\abbr{p}olarization-\abbr{d}ivision \abbr{m}ultiplexing}
\NewA{qam}{QAM}{\abbr{q}uadrature \abbr{a}mplitude \abbr{m}odulation}
\NewA{sqp}{SQP}{\abbr{s}equential \abbr{q}uadratic \abbr{p}rogramming}
\NewA{rd}{RD}{\abbr{r}adii \abbr{d}istribution}
\NewA{sep}{SEP}{\abbr{s}ymbol \abbr{e}rror \abbr{p}robability}
\NewA{ser}{SER}{\abbr{s}ymbol \abbr{e}rror \abbr{r}ate}
\NewA{si}{SI}{\abbr{s}ide \abbr{i}nformation}
\NewA{sp}{SP}{\abbr{s}ingle-\abbr{p}olarization}
\NewA{sanr}{SNR}{\abbr{s}ignal-to-(additive-)\abbr{n}oise \abbr{r}atio}
\NewA{snr}{SNR}{\abbr{s}ignal-to-\abbr{n}oise \abbr{r}atio}
\NewA{snlse}{sNLSE}{\abbr{s}tochastic \abbr{n}onlinear \abbr{S}chr\"odinger \abbr{e}quation}
\NewA{nlse}{NLSE}{\abbr{n}onlinear \abbr{S}chr\"odinger \abbr{e}quation}
\NewA{stwrc}{sTRC}{\abbr{s}eparated \abbr{t}wo-way \abbr{r}elay \abbr{c}hannel}
\NewA{stwtrc}{sTTRC}{\abbr{s}eparated \abbr{t}wo-way \abbr{t}wo-\abbr{r}elay \abbr{c}hannel}
\NewA{ts}{TS}{\abbr{t}wo-\abbr{s}tage}
\NewA{twrc}{TRC}{\abbr{t}wo-way \abbr{r}elay \abbr{c}hannel}
\NewA{twtrc}{TTRC}{\abbr{t}wo-way \abbr{t}wo-\abbr{r}elay \abbr{c}hannel}
\NewA{wdm}{WDM}{\abbr{w}avelength-\abbr{d}ivision \abbr{m}ultiplexing}
\NewA{vn}{VN}{variable node}
\NewA{cn}{CN}{constraint node}
\NewA{de}{DE}{density evolution}
\NewA{sc}{SC}{spatially-coupled}
\NewA{ldpc}{LDPC}{low-density parity-check}
\NewA{bp}{BP}{belief propagation}
\NewA{dm}{DM}{dispersion-managed}
\NewA{spm}{SPM}{self-phase modulation}
\NewA{xpm}{XPM}{cross-phase modulation}
\NewA{fwm}{FWM}{four-wave-mixing}
\NewA{ixpm}{IXPM}{intrachannel cross-phase modulation}
\NewA{ifwm}{IFWM}{intrachannel four-wave-mixing}
\NewA{ssfm}{SSFM}{split-step Fourier method}
\NewA{fec}{FEC}{forward error correction}
\NewA{psd}{PSD}{power spectral density}
\NewA{ook}{OOK}{on-off keying}
\NewA{smf}{SMF}{single-mode fiber}
\NewA{ssmf}{SSMF}{standard single-mode fiber}
\NewA{dbp}{DBP}{digital backpropagation}
\NewA{edfa}{EDFA}{erbium-doped fiber amplifier}
\NewA{ofdm}{OFDM}{orthogonal frequency division multiplexing}
\NewA{exit}{EXIT}{extrinsic information transfer}
\NewA{pexit}{P-EXIT}{protograph extrinsic information transfer}
\NewA{osnr}{OSNR}{optical signal-to-noise ratio}
\NewA{roadm}{ROADM}{reconfigurable optical add-drop multiplexer}
\NewA{rps}{RPS}{Raman pump station}
\NewA{mlse}{MLSE}{maximum likelihood sequence estimation}
\NewA{dcf}{DCF}{dispersion compensating fiber}
\NewA{tcm}{TCM}{trellis coded modulation}
\NewA{edc}{EDC}{electronic dispersion compensation}
\NewA{ldpcc}{LDPCC}{low-density parity-check convolutional}
\NewA{llr}{LLR}{log-likelihood ratio}
\NewA{ra}{RA}{repeat-accumulate}
\NewA{ira}{IRA}{irregular-repeat-accumulate}
\NewA{ara}{ARA}{accumulate-repeat-accumulate}
\NewA{mi}{MI}{mutual information}
\NewA{vdmm}{VDMM}{variable degree matched mapping}
\NewA{gmi}{GMI}{generalized mutual information}
\NewA{wd}{WD}{windowed decoder}
\NewA{gn}{GN}{Gaussian noise}
\NewA{hd}{HD}{hard-decision}
\NewA{hdd}{HDD}{hard-decision decoding}
\NewA{sd}{SD}{soft-decision}
\NewA{sdd}{SDD}{soft-decision decoding}
\NewA{emp}{EMP}{extrinsic message passing}
\NewA{imp}{IMP}{intrinsic message passing}
\NewA{gldpc}{GLDPC}{generalized low-density parity-check}
\NewA{scgldpc}{SC-GLDPC}{spatially-coupled generalized low-density parity-check}
\NewA{scldpc}{SC-LDPC}{spatially-coupled low-density parity-check}
\NewA{scfec}{SC-FEC}{spatially-coupled forward error correction}
\NewA{tbbc}{TBBC}{tightly-braided block code}
\NewA{gpc}{GPC}{generalized product code}
\NewA{otn}{OTN}{optical transport network}
\NewA{bch}{BCH}{Bose--Chaudhuri--Hocquenghem}
\NewA{rs}{RS}{Reed--Solomon}
\NewA{bbbc}{BBBC}{block-wise braided block code}
\NewA{hpc}{HPC}{half-product code}
\NewA{pc}{PC}{product code}
\NewA{rv}{RV}{random variable}
\NewA{mbp}{MBP}{multiple block permutator}
\NewA{cer}{CER}{codeword error probability}
\NewA{oh}{OH}{overhead}
\NewA{bdd}{BDD}{bounded-distance decoding}
\NewA{ncg}{NCG}{net coding gain}
\NewA{gwbp}{Galton--Watson branching process}{Galton--Watson branching process}

\newacronym[%
	longplural={binary erasure channels},%
	shortplural={BECs}%
]{bec}{BEC}{binary erasure channel}%

\begin{document}

%\title{What Can We Discover about Communications \\ from Machine Learning?}

\title{What Can Machine Learning Teach Us \\ about Communications?}


\author{
	\IEEEauthorblockN{
	Mengke Lian\IEEEauthorrefmark{1}, Christian H\"{a}ger\IEEEauthorrefmark{1}\IEEEauthorrefmark{2}, %Fabrizio Carpi\IEEEauthorrefmark{3}
	 and
	Henry D.~Pfister\IEEEauthorrefmark{1}
	\thanks{%The authors can be contacted at: mengke.lian@duke.edu, christian.haeger@chalmers.se, fabrizio.carpi@duke.edu, and	henry.pfister@duke.edu.
	This work is part of a project that has received funding
	from the European Union's Horizon 2020 research and innovation
	programme under the Marie Sk\l{}odowska-Curie grant 
	No.~749798. The work was also supported by the National
	Science Foundation (NSF) under grant No.~1609327. Any opinions,
	findings, recommendations, and conclusions expressed in this
	material are those of the authors and do not necessarily reflect the
	views of these sponsors.}}

	\IEEEauthorblockA{\IEEEauthorrefmark{1}%
	Department of Electrical and Computer Engineering, Duke University,
	Durham, North Carolina }
	%{\footnotesize \emph{(henry.pfister@duke.edu)}}
	%{(christian.haeger@chalmers.se, henry.pfister@duke.edu)}
    \IEEEauthorblockA{\IEEEauthorrefmark{2}%
	Department of Electrical Engineering,
	Chalmers University of Technology,
	%41296 
	Gothenburg, Sweden }
    %\IEEEauthorblockA{\IEEEauthorrefmark{3}Department of Information
    %Engineering, University of Parma, Parma, Italy}	
}

\maketitle

\begin{abstract}
% CH: I comment this out because it is mentioned 3x in abstract and introduction
%This paper discusses two applications of machine-learning techniques to communication systems.
Rapid improvements in machine learning over the past decade are beginning to have far-reaching effects.
For communications, engineers with limited domain expertise can now use off-the-shelf learning packages to design high-performance
systems based on simulations.
Prior to the current revolution in machine learning, the majority of communication engineers were quite aware that system parameters (such as filter coefficients) could be learned
using stochastic gradient descent.
It was not at all clear, however, that more complicated parts of the system architecture could be learned as well.

In this paper, we discuss the application of machine-learning techniques to two communications problems and focus on what can be learned from the resulting systems.
We were pleasantly surprised that the observed gains in one example have a simple explanation that only became clear in hindsight.
In essence, deep learning discovered a simple and effective strategy that had not been considered earlier.
\end{abstract}

% Redefine all acronyms that have been defined in the introduction
\glsresetall

\section{Introduction}

This paper outlines a few applications of machine-learning techniques
to communication systems and focuses on what can be learned from the
resulting systems.  First, we consider the parameterized
belief-propagation (BP) decoding of parity-check codes which was
introduced by Nachmani et al.\ in~\cite{Nachmani2016}. Then, we study
the low-complexity channel inversion known as digital backpropagation
(DBP) for optical fiber communications \cite{Ip2008}.

\section{Machine Learning}
\label{sec:machine_learning}

Before discussing the two applications in detail in
Secs.~\ref{sec:coding} and \ref{sec:optics}, we start in this section by briefly
reviewing the standard supervised learning setup for feed-forward
neural networks.  Afterwards, we highlight a few important aspects
when applying machine learning to communications problems.

\subsection{Supervised Learning of Neural Networks}

A deep feed-forward NN with $m$ layers defines a mapping $\vect{y} = f(\vect{x};\theta)$ where the input vector $\vect{x}=\vect{x}^{(0)} \in \mathcal{X}$ is mapped to the output
vector $\vect{y}=\vect{x}^{(m)} \in \mathcal{Y}$ by alternating between affine transformations (defined by $\vect{z}^{(i)} = \vect{W}^{(i)} \vect{x}^{(i-1)} + \vect{b}^{(i)}$) and
pointwise nonlinearities (defined by $\vect{x}^{(i)} = \phi(\vect{z}^{(i)})$)~\cite{LeCun2015}.
This is illustrated in the bottom part of Fig.~\ref{fig:SSFM}.
The parameter vector $\theta$ encapsulates all elements in the weight matrices $\vect{W}^{(1)}, \dots, \vect{W}^{(m)}$ and all elements in the bias vectors
$\vect{b}^{(1)},\dots,\vect{b}^{(m)}$.
Common choices for the nonlinearities include $\phi(z)=\max\{0,z\}$, $\phi(z)=\tanh(z)$, $\phi(z)=1/(1+e^{-z})$.

In a supervised learning setting, one has a training set $S\subset
\mathcal{X} \times \mathcal{Y}$ containing a list of desired
$(\vect{x},\vect{y})$ input--output  pairs.
Then, training proceeds by minimizing the empirical training loss
$\mathcal{L}_S (\theta)$, where the empirical loss $\mathcal{L}_A
(\theta)$ for a finite set $A\subset \mathcal{X}\times \mathcal{Y}$ of
input--output pairs is defined by
\[ \mathcal{L}_A (\theta) \triangleq \frac{1}{|A|} \sum_{(\vect{x},\vect{y})\in A} L \big( f(\vect{x};\theta),\vect{y}) \big) \]
and $L(\hat{\vect{y}},\vect{y})$ is the loss associated with returning the output $\hat{\vect{y}}$ when $\vect{y}$ is correct.
When the training set is large, one typically chooses the parameter
vector $\theta$ using a variant of stochastic gradient descent (SGD).
In particular, mini-batch SGD uses the parameter update
\[ \theta_{t+1} = \theta_t - \alpha \nabla \mathcal{L}_{B_t} (\theta_t), \] % + \beta (\theta_t - \theta_{t-1}), \]
where $\alpha$ is the step size and $B_t \subseteq S$ is the mini-batch used by the $t$-th step.
Typically, $B_t$ is chosen to be a random subset of $S$ with some fixed size (e.g., $|B_t| = 64$) that matches available computational resources (e.g., GPUs).

\iffalse
Until 2006, it was widely believed that deep neural networks could not be trained effectively in practice.
Since then, multiple significant advances (both in algorithms and hardware) have enabled the effective training of very deep NNs~[cite???].
Additionally, their ability perform well on test data not used during training (i.e., to generalize) is much better than one would expect given their large number of parameters~[cite???].
A variety of explanations have been suggested for this exceptional behavior.
A strong contender is that training with SGD introduces an implicit regularization of the problem that favors large-volume subsets of the parameter space containing a large fraction of good weights.
In contrast, standard gradient descent may find a parameter vector that is delicately tuned to give good performance on the training set but not robust to small perturbations.
In practice, post-2014 regularization techniques such as dropout and batch normalization can also be helpful when training deep NNs.
\fi

\subsection{Machine Learning for Communications}

Machine learning for communications differs from traditional machine
learning in a number of ways.

\subsubsection{Accurate generative modeling and infinite training data
supply}

Machine learning is typically applied to fixed-size data sets, which
are split into training and test sets. A central problem in this case
is the \emph{generalization error} caused by overfitting the model
parameters to peculiarities in the training set. On the other hand,
communication theory traditionally assumes that one can accurately
simulate and/or model the communication channel. In this case, one can
generate an infinite supply of training data with which to learn.

%However, for a particular problem, the required amount of training
%data is still unclear.

\subsubsection{Exponential number of classes}

For classification tasks, a different type of generalization error is
caused by a lack of class diversity in the training set. For classical
machine learning applications, there are typically only few classes
and the training set contains a sufficient number of training examples
per class. On the other hand, for certain communications problems,
e.g., decoding error-correcting codes, the number of classes increases
exponentially with the problem size. Training unrestricted NNs (even
deep ones) with only a subset of classes leads to poor generalization
performance \cite{Gruber2017}. 

% Due to the massive number of distinct messages and
%noise patterns, it is clear that some notion of generalization will be
%required for good performance. 

%In~\cite{Gruber2017}, it is shown that unrestricted NNs of the same
%complexity perform better with linear codes than random codes.  

\subsubsection{Black-box computation graphs vs.~domain knowledge}

Another consequence of having an accurate channel model is that one
can actually implement optimal or close-to-optimal solutions in many
cases. In that case, learning can be motivated as a means to reduce
complexity because there may exist simple approximations with much
lower complexity. Moreover, existing domain knowledge can be used to
simplify the learning task. Indeed, for both considered applications
in this paper, one actually improves existing algorithms by extensively
parameterizing their associated computation graphs, rather than
optimizing conventional ``black-box'' NN architectures. Our focus is
on examining the trained solutions and trying to understand why they
work better and solve the problem more efficiently than the hand-tuned
algorithms they are based on.

\section{Optimized BP Decoding of Codes}
\label{sec:coding}

%\footnote{We remark that there are
%less-restrictive NN decoder designs that also take advantage of code
%linearity~\cite{Tallini1995,Bennatan2018}.}

Recently, Nachmani, Be'ery, and Burshtein proposed a weighted BP (WBP)
decoder with different weights (or scale factors) for each edge in the
Tanner graph ~\cite{Nachmani2016}. These weights are then optimized
empirically using tools and software from deep learning. One of the
main advantages of this approach is that the decoder automatically
respects both code and channel symmetry and requires many fewer
training patterns to learn. Their results show that this approach
provides moderate gains over standard BP when applied to the
parity-check matrices of BCH codes. A more comprehensive treatment of
this idea can be found in~\cite{Nachmani2018}. In addition, there are other
less-restrictive NN decoders that also take advantage of code and channel
symmetry~\cite{Tallini1995,Bennatan2018}

%A neural network (NN) decoder for error-correcting codes that respects
%both code and channel symmetry (e.g., linear codes on symmetric
%channels) should require many fewer training patterns to learn. For
%example, in~\cite{Nachmani2016}, the NN is restricted to a BP-like
%decoder that naturally takes advantage of these properties and
%low-weight parity checks. Finally, there are less-restrictive NN
%designs that also take advantage of
%linearity~\cite{Tallini1995,Bennatan2018}.

While the performance gains of WBP decoding are worth investigating, the additional complexity of storing and applying one weight per edge is significant.
In our experiments, we also consider \emph{simple scaling} models that share weights to reduce the storage and computational burden.
In these models, three scalar parameters are used for each iteration: the message scaling, the channel scaling, and the damping factor.
They can also be shared for all iterations.

%!TEX root = ../tex/paper.tex
% Single braces
\newcommand{\rdbrs}[1]{\left( #1 \right)}               % round braces,     ( x )
\newcommand{\sqbrs}[1]{\left\lbrack #1 \right\rbrack}   % square braces,    [ x ]
\newcommand{\clbrs}[1]{\left\lbrace #1 \right\rbrace}   % curly braces,     { x }

% Delimited braces
\newcommand{\rdbrsv}[2]{\left( #1 \,\middle|\, #2 \right)}              % round braces with delimiter |,    ( x | y ), conditional probability
\newcommand{\sqbrsv}[2]{\left\lbrack #1 \,\middle|\, #2 \right\rbrack}  % square braces with delimiter |,   [ x | y ], conditional expectation
\newcommand{\clbrsv}[2]{\left\lbrace #1 \,\middle|\, #2 \right\rbrace}  % curly braces with delimiter |,    { x | y }, set

\newcommand{\abs}[1]{\left| #1 \right|}     % Absolute value,   | x |

% Short for boldsymbol
\newcommand{\bs}[1]{\boldsymbol{#1}}

% three different levels of decoder complexity: simple scaling, 
% ich are generally irregular (e.g., different rows ()rows and columns may have different number)
% Henry/Mengke: Describe recent work
% Complexity
% symmetry
% simple scaling
% damping
% performance
% overcomplete parity check matrices
% RRD

\subsection{Weighted Belief-Propagation Decoding}
% brief synopsis + citations. Only equations necessary for later descriptions.
Consider an $ [N, K] $ linear code $ \mathcal{C} $ defined by an $ M \times N $ parity-check matrix $H$.
Given any parity-check matrix $H$, one can construct a bipartite
Tanner graph $\mathcal{G}=(V,C,E)$, where $ V \triangleq [N]$ and $C \triangleq [M] $ are sets of variable nodes (i.e., code symbols) and check nodes (i.e., parity constraints).
The edges, $ E = \clbrsv{ (v,c) \in V \times C }{ H_{cv} \ne 0 } $, connect all parity checks to the variables involved in them.
By convention, the boundary symbol $ \partial $ denotes the neighborhood operator defined by
\begin{equation*}
    \partial v \triangleq \clbrsv{ c }{ (v,c) \in E }, \quad \partial c \triangleq \clbrsv{ v }{ (v,c) \in E }.
\end{equation*}

The log-likelihood ratio (LLR) is the standard message representation for BP decoding of binary variables.
The initial channel LLR for variable node $ v \in V $ is defined by
\begin{equation} \label{eq:LLR_def}
    \ell_v \triangleq \log \rdbrs{ \frac{ \mathrm{Pr} \rdbrsv{ y_v }{ x_v = 0 } }{ \mathrm{Pr} \rdbrsv{ y_v }{ x_v = 1 } } },
\end{equation}
where $ y_v $ is the $ v $-th symbol in the channel output sequence, and
$ x_v $ is the corresponding bit in the transmitted codeword.

WBP is an iterative algorithm that passes messages along the edges of the Tanner graph $ \mathcal{G} $.
During the $t$-th iteration, a pair of messages $ \hat{\lambda}_{c \to v}^{(t)} $, and $ \lambda_{v \to c}^{(t)} $ are passed in each direction along the edge $ (v,c) $.
This occurs in two steps: the check-to-variable step updates messages $ \bs{\hat{\lambda}}^{(t)} \triangleq \{ \hat{\lambda}_{c \to v}^{(t)} \}_{(v,c) \in E} $ and variable-to-check step updates messages $ \bs{\lambda}^{(t)} \triangleq \{ \lambda_{v \to c}^{(t)} \}_{(v,c) \in E} $.
In the variable-to-check step, the pre-update rule is
\begin{equation} \label{eq:LBP_vertical}
    \lambda_{v \to c}^{\prime (t)}
    = w_v^{(t)} \ell_v + \sum_{c^\prime \in \partial v \setminus c} w_{v^\prime c}^{(t)} \hat{\lambda}_{c^\prime \to v}^{(t-1)}, 
\end{equation}
where $ w_{vc}^{(t)} $ is the weight assigned to the edge $(v,c)$ and $ w_v^{(t)} $ is the weight assigned to the channel LLR $ \ell_v $.
In the check-to-variable step, the pre-update rule is
\begin{equation} \label{eq:LBP_horizontal}
    \hat{\lambda}_{c \to v}^{\prime (t)} 
    = 2 \tanh^{-1} \rdbrs{ \prod_{v^\prime \in \partial c \setminus v} \tanh \rdbrs{ \frac{\lambda_{v^\prime \to c}^{(t)}}{2} } }.
\end{equation}
To avoid numerical issues, the absolute value of $\hat{\lambda}_{c \to v}^{\prime (t)} $ is clipped, if it is larger than some fixed value (e.g., 15).

To mitigate oscillation and enhance convergence, we also use a \emph{damping} coefficient $ \gamma \in [0, 1] $ to complete the message updates~\cite{Fossorier-istc03}.
Damping is referred to as ``relaxed BP'' in
\cite{Nachmani2018}, where it is studied in the context of weighted
min-sum decoding. 
This method of improving performance was not considered in~\cite{Nachmani2016}.
In particular, the final BP messages at iteration $ t $ are computed using a convex combination of the previous value and the pre-update value:
\begin{IEEEeqnarray}{rCl}
    \lambda_{v \to c}^{(t)} &=& (1-\gamma) \lambda_{v \to c}^{(t-1)} + \gamma \lambda_{v \to c}^{\prime (t)}, \label{eq:damping_vertical} \\
    \hat{\lambda}_{c \to v}^{(t)} &=& (1-\gamma) \hat{\lambda}_{c \to v}^{(t-1)} + \gamma \hat{\lambda}_{c \to v}^{\prime (t)}. \label{eq:damping_horizontal}
\end{IEEEeqnarray}

For the marginalization step, the sigmoid function $ \sigma(x) \equiv \rdbrs{ 1 + e^{-x} }^{-1} $ is used to map the output LLR to an estimate $o_v^{(t)}$ of the probability that $ x_v = 0 $ defined by
\begin{equation} \label{eq:LBP_margin}
    o_v^{(t)} = \sigma \Bigl( w_v^{(t)} \ell_v + \sum_{v^\prime \in \partial c} w_{v^\prime c}^{(t)} \hat{\lambda}_{c \to v^\prime}^{(t)} \Bigr).
\end{equation}
Setting all weights to $1$ and $\gamma = 1$ recovers standard BP.

\subsection{From WBP to Optimized WBP}
Any iterative algorithm, such as WBP decoding, can be ``unrolled'' to give a feed-forward architecture that has the same output for some fixed number of iterations~\cite{Gregor-icml10}.
Moreover, the sections in the feed-forward architecture are not required to be identical.
This increases the number ``trainable'' parameters that can be optimized.

It is well-known that BP performs exact marginalization when the Tanner graph is a tree, but good codes typically have loopy Tanner graph with short cycles.
To improve the BP performance on short high-density parity-check (HDPC) codes, one can optimize the weights $ w_{vc}^{(t)} $ and $ w_{v}^{(t)} $ in all iterations ~\cite{Nachmani2016}.
The damping coefficient $\gamma$ can also be optimized.

%\input{../mengke_part/learned_BP_graph}
%In Figure \mbox{\ref{fig:learned_BP_architechure}}, each column corresponds to an entire BP iteration.

For supervised classification problems, one typically uses the cross-entropy loss function, and this loss function has also been proposed for the optimized WBP decoding problem~\cite{Nachmani2016}.
However, our experiments show that minimizing this loss may not actually minimize the bit error rate.
Instead, we use the modified loss function
\begin{equation} \label{eq:single_loss}
	L_s (\mathbf{o}^{(T)}, \mathbf{x}) \triangleq \frac{1}{N} \sum_{v=1}^N \left[ 1 + \left( \frac{o_v^{(T)}}{1-o_v^{(T)}} \right)^{1-2x_v} \right]^{-1},
\end{equation}
where $T$ is the total number of iterations. 
More details about the modified loss can be found in~\cite{Lian-isit19}.
Our experiments also show that the optimization behaves better with the multi-loss approach proposed by~\cite{Nachmani2016}.
Thus, the results in this paper are based on optimizing the modified multi-loss function
\begin{equation} \label{eq:multi_loss}
    L( \{\mathbf{o}^{(t)}\}_{t=1}^T, \mathbf{x}) \triangleq \frac{1}{T} \sum_{t=1}^{T} L_s (\mathbf{o}^{(t)}, \mathbf{x}).
\end{equation}

\iffalse
The training aims at minimizing the cross entropy loss function
\begin{equation} \label{eq:single_loss}
   L(\mathbf{o}, \mathbf{x}) \triangleq -\frac{1}{N} \sum_{v=1}^N \sqbrs{ x_v \log(1-o_v) \! +\! (1-x_v) \log(o_v) }.
\end{equation}

To improve learning, \cite{Nachmani2016} also suggests to utilize the marginalization at each iteration to define a multi-loss variant of \eqref{eq:single_loss}:
\begin{equation} \label{eq:multi_loss}
    L_{\mathrm{multi}}( \{\mathbf{o}^{(t)}\}_{t=1}^T, \mathbf{x}) = \frac{1}{T} \sum_{t=1}^{T} L(\mathbf{o}^{(t)}, \mathbf{x}).
\end{equation}
\fi

The optimization complexity depends on the number of iterations and how the parameters are shared.
For example, one can share the weights temporally (across decoding iterations) and/or spatially (across edges):
\begin{itemize}
\item   If the weights are shared temporally, i.e.,
        \begin{equation*}
             w_v^{(t)} \equiv w_v, \qquad w_{vc}^{(t)} \equiv w_{vc}, \qquad \forall\, t ,
        \end{equation*}
        one obtains a recursive NN (RNN) structure.
\item   If the weights are shared spatially, i.e.,
        \begin{equation*}
            w_v^{(t)} \equiv w_{\mathrm{ch}}^{(t)}, \qquad w_{vc}^{(t)} \equiv w_{\mathrm{msg}}^{(t)}, \qquad \forall\, (v, c) \in E,
        \end{equation*}
        then there are only two scalar parameters per iteration: one for the channel LLR and one for the BP messages.
        Compared to the fully weighted (FW) decoder, we call this the simple scaled (SS) decoder.
        
\item Sharing weights both temporally and spatially results in only two weight parameters, $w_{\mathrm{ch}}$ and $w_{\mathrm{msg}}$. 
\end{itemize}

\subsection{Random Redundant Decoding (RRD)}

A straightforward way to improve BP decoding for HDPC codes is to use redundant parity checks (e.g., by adding dual codewords as rows to the parity-check matrix)~\cite{Yedidia-aller02}.
In general, however, the complexity of BP decoding scales linearly with the number of rows in the parity-check matrix.

Another approach is to spread these different parity checks over time, i.e., by using different parity-check matrices in each iteration~\cite{Halford-isit06,Dimnik-com09,Hehn-com10}.
This can be implemented efficiently by exploiting the code's automorphism group and reordering the code bits after each iteration in a way that effectively uses many different parity-check matrices but stores only one.

In~\cite{Nachmani2018}, optimized weighted RRD decoders are constructed by cascading several WBP blocks and reordering the code bits after each WBP block.
In this work, we also consider optimized RRD decoding based on their approach.
But, the input to $ (\tau+1) $-th learned BP block is modified to be a weighted convex combination between the initial channel LLRs and the output of the $ \tau $-th learned BP.
This procedure is similar to damping and the \emph{mixing} coefficient $ \beta $ is also learned.

For RRD decoding, choosing a good parity-check matrix is crucial because the code automorphisms permute the variable nodes without changing the structure of the Tanner graph.
In general, good Tanner graphs have fewer short cycles and can be constructed with heuristic cycle-reduction algorithms \cite{Halford-isit06}.
%For cod or by randomly choosing parity-check rows to be minimum-weight dual codewords.

%\input{../mengke_part/learned_BP_RRD_graph}

\begin{figure}[t]
    \centering
    \scalebox{1.05}{\includegraphics{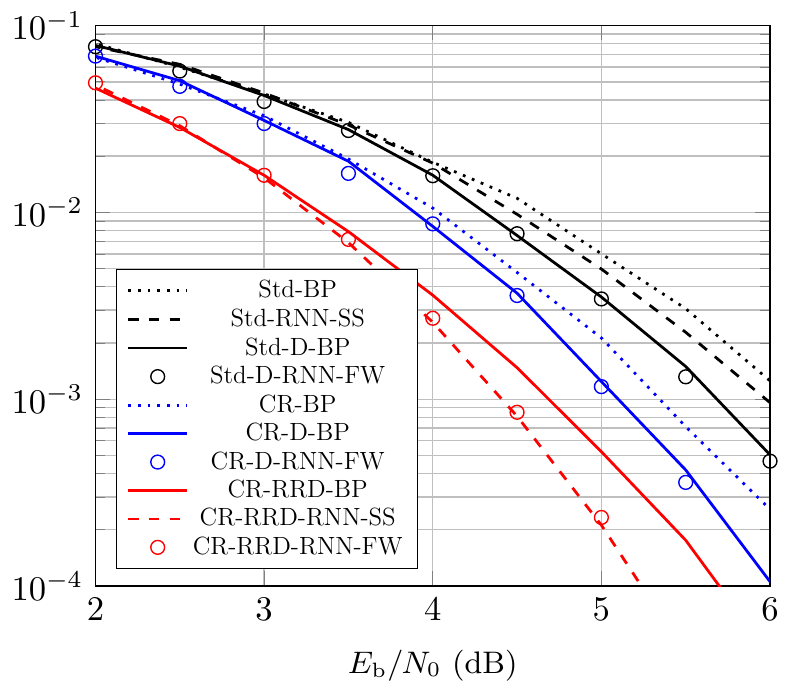}}
    \caption{BER results for the $ \mathsf{BCH}(63, 36) $ code. Curves are labeled to indicate: whether the parity-check matrix is standard (Std) or cycle reduced (CR), whether damping (D) is used, and also to show the decoder style (BP/RNN-SS/RNN-FW).  }
    \label{fig:BCH_63_36_curves}
    \vspace{-1mm}
\end{figure}

\subsection{Experiments and  Results}

The various feed-forward decoding architectures in this paper are implemented in the PyTorch framework and optimized using the RMSPROP optimizer. %\cite{Tieleman2012}.
The number of decoding iterations is set to $T = 20$.
For the RRD algorithm, the code bits are permuted after every second decoding iteration
and optimized (iteration-independent) mixing and damping coefficients are used.
The decoder architectures are trained using transmit-receive pairs for the binary-input AWGN channel where the $E_b/N_0$ SNR parameters is chosen uniformly between $ 1 $ dB and $ 6 $ dB for each training pair.
To avoid numerical issues, the gradient clipping threshold is set to $ 0.1 $ and the LLR clipping threshold is $ 15 $.
We define each epoch to be $ 1000 $ mini-batches and each mini-batch to be $ 100 $ transmit-receive pairs.
All decoders are trained for $ 20 $ epochs and optimized using the multi-loss function \eqref{eq:multi_loss}.

In Fig.~\ref{fig:BCH_63_36_curves}, we show the performance curves achieved by the optimized decoders for the $ \mathsf{BCH}(63, 36) $ code.
For the standard parity-check matrix without RRD, the standard BP decoder with damping (Std-D-BP) performs very similarly to the FW optimized decoder (Std-D-RNN-FW).
Similarly, for the cycle-reduced parity-check matrix, damping (CR-D-BP) achieves essentially the same gain as the fully-weighted model (CR-S-RNN-FW).
Thus, the dominant effects are fully explained by using damping and cycle-reduced parity-check matrices.

For a similar complexity, the RRD algorithm achieves better results.
This is true both for standard BP (CR-RRD-BP) with optimized mixing and damping and for optimized weights (CR-RRD-RNN-SS) in the simple-scaling model.
However, the fully-weighted model (CR-RRD-RNN-FW) does not provide significant gains over simple scaling.
Also, RRD results are shown only for cycle-reduced matrices because they perform much better.

\iffalse
\subsection{Effect of Symmetry}

The work in~\cite{Nachmani2016} uses standard parity-check matrices for BCH codes while the follow-up work in~\cite{Nachmani2018} considers cycle-reduced parity-check matrices and RRD.
In all cases, the rows (and columns) of these matrices typically have different Hamming weights from each other.
This is one reason why one might expect optimized WBP to outperform standard BP.

To test this hypothesis, we also considered optimized WBP for parity-check matrices with more symmetry.
For cyclic codes, one can construct a square parity-check matrix whose rows are given by cyclically shifting a single parity check.
For Reed--Muller codes, one can form a redundant parity-check matrix whose rows enumerate the set of minimum-weight parity-checks.
For these cases, our experiments show the performance of the simple-scaled model matches the performance of the fully-weighted model very closely~\cite{Lian-isit19}.
\fi

\section{Machine Learning for Fiber-Optic Systems}
\label{sec:optics}

In this section, we discuss the application of machine learning
techniques to optical-fiber communications. 

%The per-channel data rates in these types of systems are in the order
%of hundreds of Gbit/s. Providing reliable communication at such high
%data rates is a signficant challenge. 
%
%Fiber-optic communication systems enable high-speed data traffic over
%very long distances. 
%
%In a single-mode optical fiber, narrowband signals propagate according
%to the nonlinear Schr\"odinger equation (NLSE)
%\cite[p.~40]{Agrawal2006}. 

%Our design choices are directly motivated by such considerations. In
%particular, our computation graph exploits the hierarchical problem
%structure that is introduced by the transmission process. Moreover, we
%choose the linear operators in the graph to be short and symmetric
%finite impulse response (FIR) filters. 

%\cite{Essiambre2005, Li2008, Ip2008}
%, Roberts2006

% \cite{IpKahn2009}
% filtering: \cite{Li2005a} ? 

% Mateo2008 talks about impact on inter-channel interference
% Li2008 talks about filtering and WDM

%, Li2011
%,Nakashima2017a

%Tao2011,

\subsection{Signal Propagation and Digital Backpropagation}

%and was inspired by a similar idea where optical components were used
%for the processing \cite{Pare1996}. 
%DBP was first studied as a transmitter pre-distortion technique
%\cite{Essiambre2005, Roberts2006}.

\begin{figure}[t]
	\scalebox{0.93}{\includegraphics{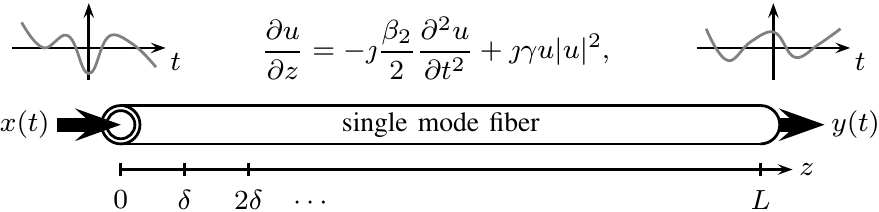}}
	%\vspace*{-0.2cm}
	\caption{Conceptual signal evolution in a single-mode fiber. The
	nonlinear Schr\"odinger equation implicitly describes the
	relationship between the input signal $x(t) = u(z=0,t)$ and the
	output signal $y(t) = u(z=L,t)$. The parameters $\beta_2$ and
	$\gamma$ are, respectively, the chromatic dispersion coefficient and
	the nonlinear Kerr parameter. }
	\label{fig:fiber}
\end{figure}

Fiber-optic communication links carry virtually all intercontinental
data traffic and are often referred to as the \emph{Internet
backbone}. We consider a simple point-to-point scenario, where a
signal with complex baseband representation $x(t)$ is launched into an
optical fiber as illustrated in Fig.~\ref{fig:fiber}.  The signal
evolution is implicitly described by the nonlinear Schr\"odinger
equation (NLSE) which captures dispersive and nonlinear propagation
impairments~\cite[p.~40]{Agrawal2006}.
 After distance $z=L$, the
received signal $y(t)$ is low-pass filtered and sampled at $t = k T$
to give the samples $\{y_k\}_{k \in \mathbb{Z}}$. 

%The loss term $\alpha u/2$, where $\alpha$ is the attenuation
%parameter, is ignored for simplicity.

%Consider
%now the time-discretized NLSE
%\footnote{Eq.~\eqref{eq:discretized_nlse} can be obtained from the
%NLSE by discretizing the differential operator
%$\frac{\partial^2}{\partial t^2}$ in the spectral domain and applying
%an inverse DFT. Periodic boundary conditions for $x(t)$ are assumed.
%}
%\begin{align}
%	\label{eq:discretized_nlse}
%	\frac{\D \vect{u}(z)}{\D z} 
%	= \mat{A} \vect{u}(z) +
%	\imag \gamma \bm{\rho}(\vect{u}(z)), %= \vect{f}(\vect{u}(z))
%\end{align}
%where $\vect{A} = \FFTm^{-1} \diag(H_1, \dots, H_n) \FFTm$, $\FFTm$ is
%the $n\times n$ discrete Fourier transform (DFT) matrix, $H_k = -\imag
%\frac{\beta_2}{2} \omega_k^2$, $\omega_k = 2 \pi f_k$ is the $k$-th
%DFT angular frequency, and $\bm{\rho}$ is defined as the element-wise
%application of $\rho(x) = x |x|^2$. 

% using the received samples $\{y_k\}_{k \in \mathbb{Z}}$ as a
%boundary condition

%is based on a block-wise processing where $n$
%received samples are collected into a vector $\vect{y} = (y_1, \dots,
%y_n)^\transpose \in \mathbb{C}^n$. The fiber is then 

\begin{figure}[t]
	\centering
		\scalebox{0.93}{\includegraphics{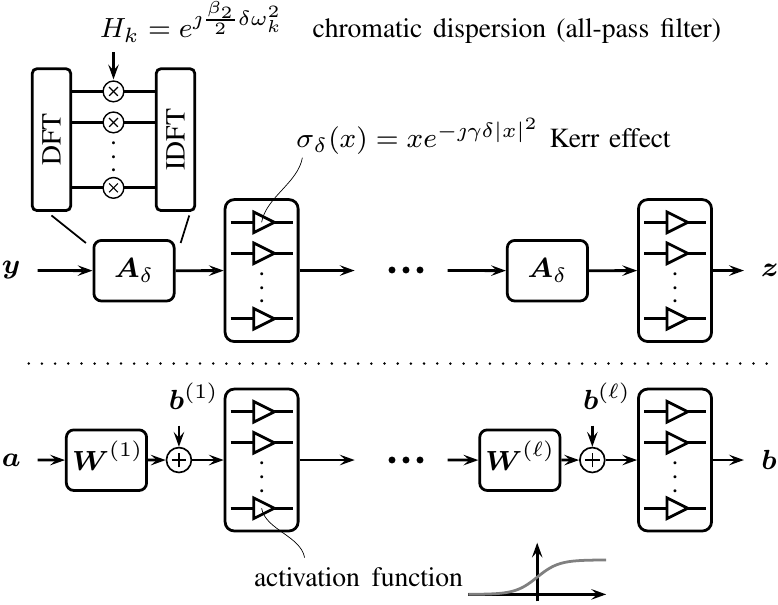}}
	\caption{Block diagram of the split-step Fourier method to
	numerically solve the nonlinear Schr\"odinger equation (top) and the
	canonical model of a deep feed-forward neural network (bottom).}
	\label{fig:SSFM}
	\vspace{-1mm}
\end{figure}

In the absence of noise, the NLSE is invertible and the transmitted
signal can be recovered by solving the NLSE in the reverse propagation
direction.  This approach is referred to as digital backpropagation
(DBP) in the literature. DBP requires a numerical method to solve the
NLSE and a widely studied method is the split-step Fourier method
(SSFM). The SSFM conceptually divides the fiber into $\NumSteps$
segments of length $\StepSize = L / \NumSteps$ and it is assumed that
for sufficiently small $\delta$, the dispersive and nonlinear effects
act independently. A block diagram of the SSFM for DBP is shown in the
top part of Fig.~\ref{fig:SSFM}, where $\vect{y} = (y_1, \dots,
y_n)^\transpose$. In particular, one alternates between a linear
operator $\mat{A}_\delta$ and the element-wise application of a
nonlinear phase-shift function $\sigma_\delta(x) = x e^{\imag \gamma
\delta |x|^2 }$. Assuming a sufficiently high sampling rate, the
obtained vector ${\vect{z}}$ converges to a sampled version of $x(t)$
as $M \to \infty$. By comparing the two computation graphs in
Fig.~\ref{fig:SSFM}, one can see that the SSFM has a naturally layered
or hierarchical Markov structure, similar to a deep feed-forward NN. 

%The intermediate feature representation after the hidden neural
%network layers correspond to complex baseband representation
%representation of the signal along the propagation path. 

%Increasing $M$ leads to a more accurate approximation, but also
%increases complexity, as discussed in the next section.  The degree
%to which the obtained vector ${\vect{z}}$ constitutes a good
%approximation of $x(t)$ is now a question of choosing $M$, $T$, and
%$n$. In practice, $1/T$ is typically an integer multiple of the baud
%rate and $n$ is chosen to minimize the overhead in overlap-and-save
%techniques for continuous data transmission.

%More precisely, for $\gamma = 0$, \eqref{eq:discretized_nlse} is
%linear with solution $\vect{u}(z) = \mat{A}_z \vect{u}_0$, where
%$\mat{A}_z \define e^{z \mat{A}}$. For $\beta_2 = 0$, the solution is
%$\vect{u}(z) = \nlop{z}(\vect{u}_0)$, where $\nlop{z}$ is the
%element-wise application of $\sigma_z(x) = x e^{\imag \gamma z |x|^2
%}$. Alternating between these two operators for $z=-\delta$ leads to
%the block diagram shown in the top part of Fig.~\ref{fig:SSFM}. 

% = \FFTm^{-1} \diag(e^{z H_1}, \dots, e^{z H_n}) \FFTm

%\begin{remark}
%	In that regard, one may argue that \eqref{eq:discretized_nlse} is a
%	``computationally inefficient'' time-discretization of the NLSE, in
%	the sense that it relates local propagation changes to all time
%	instances. A different time-discretization approach is via partial
%	discretization or finite-difference methods. Indeed,
%	finite-difference methods can be more computationally efficient than
%	the SSFM in some applications \cite[Sec.~2.4.2]{Agrawal2006}.
%	However, to the best of our knowledge, finite-difference methods
%	have not been studied for real-time DBP. One reason for this might
%	be that many methods that show good performance are implicit, i.e.,
%	they require solving a system of equations at each step. This makes
%	it challenging to satisfy a real-time constraint. 
%\end{remark}

%Ignoring the complexity of the nonlinear steps, the SSFM can be
%implemented using $M$ DFT/IDFT pairs, utilizing the fast Fourier
%transform. On the other hand, a linear equalizer can be implemented
%using a single DFT/IDFT pair.  Based on this reasoning, the SSFM is at
%least $M$ times more complex than linear equalization. This motivates
%a number of approaches that focus on reducing the number of steps,
%see, e.g., \cite{Du2010, Rafique2011a, Secondini2016} and references
%therein. 

\subsection{Parameter-Efficient Learned Digital Backpropagation}

%We approach this problem from a machine-learning perspective. Our
%approach is to obtain a multi-layer computation graph similar to a
%deep NN by applying the split-step Fourier method (SSFM)
%\cite{Agrawal2006}. This can be seen as an example of a more general
%methodology where domain knowledge is used to generate computation
%graphs with many layers, see, e.g., \cite{Gregor2010}. 

A major issue with DBP is the large computational burden associated
with a real-time implementation.  Despite significant efforts
to reduce complexity (see, e.g., \cite{Ip2008, Du2010, Rafique2011a}),
DBP based on the SSFM is not used in
any current optical system that we know. Instead, only linear equalizers are employed.
Their implementation already poses a significant challenge; with data
rates exceeding $100$ Gbit/s, linear equalization of chromatic
dispersion is typically one of the most power-hungry receiver blocks
\cite{Pillai2014}. 

Note that the linear propagation operator $\mat{A}_\delta$ in the SSFM
is a dense matrix. On the other hand, deep NNs are typically designed
to have very sparse weight matrices in most of the layers to achieve
computational efficiency.  Sparsification of $\mat{A}_\delta$ can be
achieved by switching from a frequency-domain to a time-domain
filtering approach using finite-impulse response (FIR) filters. The
main challenge in that case is to find short FIR filters in each SSFM
step that approximate well the ideal chromatic dispersion all-pass
frequency response. In previous work, the general approach is to
design either a single filter or filter pair and use is repeatedly in
each step \cite{Ip2008, Zhu2009, Goldfarb2009, Fougstedt2017}.
However, this typically leads to poor parameter efficiency (i.e., it
requires relatively long filters) because truncation errors pile up
coherently.  We have shown in \cite{Haeger2018ofc, Haeger2018isit}
that this truncation error problem can be controlled effectively by
performing a \emph{joint} optimization of all filter coefficients in
the entire DBP algorithm. In particular, the computation graph of the
SSFM is optimized via SGD by simply
interpreting all matrices $\mat{A}_\delta$ as tunable parameters
corresponding to the FIR filters, similar to the weight matrices in a
deep NN. The nonlinearities are left unchanged, i.e., they correspond
to the nonlinear phase-shift functions in the original SSFM and not to
a traditional NN activation function. The resulting method is
referred to as learned DBP (LDBP).

%In other words, the resulting parameterized multi-layer computation
%graph is obtained using domain knowledge and it does not correspond to
%a ``black-box'' function approximator. 

%such as direct sampling and truncation \cite{Savory2008},
%frequency-domain sampling \cite{Ip2008}, 
%using a wide variety of methods such as wavelets
%\cite{Goldfarb2009} or least-squares \cite{Eghbali2014, Sheikh2016},

\subsection{Optimization Results}

\begin{figure}[t]
	\centering
		\scalebox{1.15}{\includegraphics{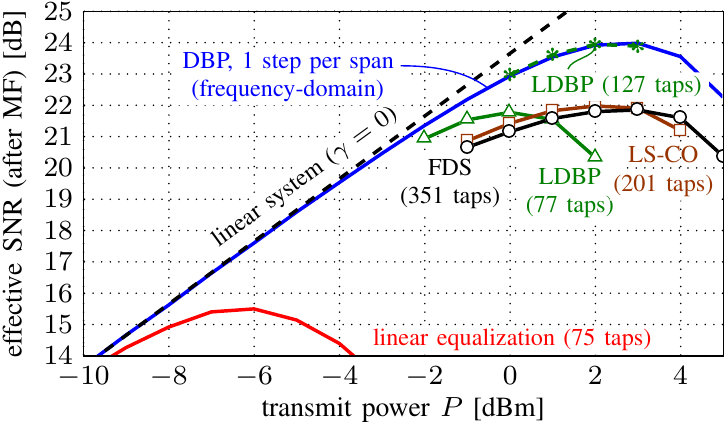}}
\vspace{-0mm}
		\caption{Results for
	$25$ spans of $80\,$km
		single-mode fiber (see parameters in
		\cite{Haeger2018isit}). FDS: frequency-domain sampling,
		LS-CO: least-squares-optimal constrained out-of-band gain, MF: matched filter. }
	\label{fig:results}
	\vspace{-1mm}
\end{figure}

In Fig.~\ref{fig:results}, we compare the equalizer accuracy in terms
of the effective SNR of LDBP to the conventional approach of designing
a single FIR filter (either via least-squares fitting or
frequency-domain sampling) and then using it repeatedly in the SSFM.
LDBP requires significantly fewer total filter taps (indicated in
brackets) to achieve similiar or better peak accuracy. The obtained
FIR filters are as short as $5$ or $3$ (symmetric) taps per step,
leading to very simple and efficient hardware implementation. This is confirmed by recent
ASIC synthesis results which show that the power consumption of LDBP becomes
comparable to linear equalization \cite{Fougstedt2018ecoc}.
LDBP can also be extended to subband processing to enable low-complexity DBP for multi-channel or other wideband transmission scenarios \cite{Haeger2018ecoc}. 

%By performing a joint filter optimization based on stochastic gradient
%descent, LDBP can significantly outperform previous FIR-filter-based
%approaches to DBP. 
%
%reduce the required number of filter taps to obtain a given equalizer
%accuracy compared to other filter design approaches. Depending on the
%system parameters, the learned filters can be as short as 5 or 3 taps
%per step, leading to very simple hardware implementations. 

%(e.g., in a least-squares sense) 

At first glance, the obtained results in Fig.~\ref{fig:results} may be
somewhat counterintuitive. Indeed, after examining the optimized
individual (per-step) filter responses in LDBP, we found that they are
generally worse approximations to the ideal chromatic dispersion
frequency response compared to filters obtained by least-squares
fitting or other methods. However, the combined response of
neighboring filters and also the overall response is better compared
to the conventional strategy of using the same filter in each
dispersion compensation stage. In fact, using the same filter many
times in series magnifies any weakness. By using
different filters at each stage, the problem is avoided and shorter
filters can achieve the same performance.

%However, and somewhat surprisingly, by sacrificing some accuracy for
%the individual responses in each step, it is possible to achieve a
%better combined response of neighboring filters and also a better
%overall response. 
%
%In the second problem, we found that deep learning discovered an
%elegant solution we had not considered. All other time-domain
%approaches to DBP used the same approximate filter for each dispersion
%compensation stage.  But, any weakness in this approximation is
%magnified by using the same filter many times in series.  By using
%different filters at each stage, the problem is avoided and shorter
%filters can achieve the same performance.

%In essence, a joint optimization sacrifices some accuracy for the
%individual frequency responses in each step, but it achieves a better
%combined response of neighboring filters and also a better overall
%response.

%\begin{figure}[t]
%	\centering
%	\subfloat[$\vect{h} \define \vect{h}^{(1)} = \dots
%	= \vect{h}^{(M)}$]{\includegraphics{response_1}}
%	\quad
%	\subfloat[$\vect{h} \define \vect{h}^{(1)} * \dots *
%	\vect{h}^{(M)}$]{\includegraphics{response_25}}
%	\caption{Schematic illustration of the truncation error when using
%	the same (or very similar) FIR filters in a split-step method, where
%	$*$ denotes convolution.}
%	\label{fig:truncation_error}
%\end{figure}

\section{Conclusion}
\label{sec:conclusion}

Recent progress in machine learning and off-the-shelf learning
packages have made it tractable to add many parameters to existing
communication algorithms and optimize. In this paper, we have
reviewed this approach with the help of two applications.

For the decoding application, our experiments support the observations in~\cite{Nachmani2016,Nachmani2018} that  optimizing parameterized BP decoders can provide meaningful gains.
In addition, we observed that many fewer parameters (e.g., damping alone) may be sufficient to achieve very similar gains.
Thus, for this general approach, it can be fruitful to also minimize the parameterization necessary to achieve the same gain~\cite{Lian-isit19}.

For the digital backpropagation application, we were pleasantly surprised that, after analyzing
the learned solution, we were able to understand why it worked so
well. In essence, deep learning discovered a simple and effective
strategy that had not been considered earlier.

%In the second problem, we found that deep learning discovered an elegant solution we had not considered.
%All other time-domain approaches to DBP used the same approximate filter for each dispersion compensation stage.
%But, any weakness in this approximation is magnified by using the same filter many times in series.
%By using different filters at each stage, the problem is avoided and shorter filters can achieve the same performance.

%We have considered the problem of reducing the complexity of DBP to
%facilitate a real-time DSP implementation. Our approach, called
%learned DBP (LDBP), is based on a multi-layer computation graph
%generated by the SSFM with many steps. Computational efficiency is
%achieved by using, in each step, very short and symmetric FIR filters
%that are jointly optimized with deep learning. Numerical results show
%that for a single-channel transmission scenario, LDBP can achieve a
%favorable performance--complexity trade-off compared to other filter
%design methods and perturbation-based ``few-step'' DBP. 

%\appendices
%\appendix
%
%\section{Proof of Theorem}
%\label{app:proofs}


%{\footnotesize
%\bibliographystyle{IEEEtran}
%\bibliography{$HOME/Dropbox/lib/bibtex/library_mendeley}
%\bibliography{WCLabrv,WCLnewbib,../bibtex/library_mendeley}
%\bibliography{../bibtex/library_mendeley}

%\bibliography{WCLabrv,WCLnewbib,biblio}

%}%

\end{document}